# Optimization of Energy-Constrained IRS-NOMA Using a Complex Circle Manifold Approach

Mahmoud AlaaEldin, *Student Member, IEEE*, Emad Alsusa, *Senior Member, IEEE*, Karim Seddik, *Senior Member, IEEE*, Mohammad Al-Jarrah, *Member, IEEE*, and Constantinos Papadias, *Fellow, IEEE*

*Abstract*—This work investigates the performance of intelligent reflective surfaces (IRSs) assisted uplink non-orthogonal multiple access (NOMA) in energy-constrained networks. Specifically, we formulate and solve two optimization problems; the first aims at minimizing the sum of users' transmit power, while the second targets maximizing the system level energy efficiency (EE). The two problems are solved by jointly optimizing the users' transmit powers and the beamforming coefficients at IRS subject to the users' individual uplink rate and transmit power constraints. A novel and low complexity algorithm is developed to optimize the IRS beamforming coefficients by optimizing the objective function over a *complex circle manifold* (CCM). To efficiently optimize the IRS phase shifts over the manifold, the optimization problem is reformulated into a feasibility expansion problem which is reduced to a max-min signal-plus-interference-ratio (SINR). Then, with the aid of a smoothing technique, the exact penalty method is applied to transform the problem from constrained to unconstrained. The proposed solution is compared against three semi-definite programming (SDP)-based benchmarks which are semi-definite relaxation (SDR), SDP-difference of convex (SDP-DC) and sequential rank-one constraint relaxation (SROCR). The results show that the manifold algorithm provides better performance than the SDP-based benchmarks, and at a much lower computational complexity for both the transmit power minimization and EE maximization problems. The results also reveal that IRS-NOMA is only superior to orthogonal multiple access (OMA) when the users' target achievable rate requirements are relatively high.

## I. Introduction

Recently, intelligent reflective surfaces (IRS) have emerged as a new technology due to their potential for improving the spectral and energy efficiency of wireless networks [1]–[5]. Generally speaking, IRS is a planar surface that consists of a large number of adjustable and low-cost passive reflecting elements which reflect incident signals after adjusting their amplitudes and phase shifts. The merit of IRS lies in its ability to *engineer* the wireless channel to realize different design needs such as signal strengthening and interference mitigation.

Mahmoud AlaaEldin, Emad Alsusa, and Mohammad Al-Jarrah are with the Electrical and Electronic Engineering Department, University of Manchester, Manchester, UK M13 9PL (e-mail: mahmoud.alaaeldin@manchester.ac.uk; e.alsusa@manchester.ac.uk; mohammad.al-jarrah@manchester.ac.uk).
Karim G. Seddik is with the Department of Electronics and Communications Engineering, American University in Cairo, Cairo, Egypt 11835 (e-mail: kseddik@aucegypt.edu).
Constantinos B. Papadias is with the Research, Technology and Innovation Network (RTIN), Alba, The American College of Greece, 153 42 Athens, Greece, and also with the Department of Electronic Systems, Aalborg University, 9220 Aalborg, Denmark (e-mail: cpapadias@acg.edu).

As such, integrating IRS in different wireless applications has been extensively studied in the literature. For example, in [6], the authors studied the application of IRS in a single cell scenario with a multi-antenna base station (BS) and single-antenna users. They developed alternation minimization algorithms to minimize the total transmit power at BS by applying a semi-definite relaxation (SDR) approach to optimize the beamforming at IRS. In [7], the authors proposed efficient algorithms to maximize the energy efficiency (EE) of IRS assisted downlink communication systems with a multi-antenna BS serving multi-antenna users. More specifically, they developed efficient algorithms based on alternation maximization, gradient descent, and sequential fractional programming to jointly optimize the transmit power allocation at BS and the phase shifts at IRS. Furthermore, in [8], a joint power and user association scheme was proposed for a multi-IRS assisted multi-BS downlink millimeter-Wave system serving multiple users. Other applications of IRS include unmanned aerial vehicles (UAV) communications [9], physical layer security for covert communications [10], physical layer network coding [11] and wireless mesh back-hauling [12].

Using IRS in energy-constrained Internet of Things (IoT) systems presents an attractive solution to address the challenge of low uplink data rates, which stem from their limited battery capacity. Since massive connectivity is another crucial requirement for IoT systems, multiple access (MA) techniques play a pivotal role in accommodating a large number of IoT devices. Existing MA techniques can be broadly categorized into two groups: orthogonal multiple access (OMA) and non-orthogonal multiple access (NOMA). The NOMA MA technique stands out as a practical choice for IoT networks, offering the ability to facilitate the access of numerous devices by allowing multiple users to simultaneously use the same spectrum [13], [14]. The merits of NOMA have made it a key technique to enhance spectral efficiency (SE) and EE of wireless communication networks [15]–[17]. Unlike OMA, NOMA can serve multiple users using the same frequency/time/code resources, and applies successive interference cancellation (SIC) at the receiver to reduce co-user interference [17], [18]. In comparison to OMA, integrating NOMA into conventional IoT applications without IRS has demonstrated several advantages, particularly in applications like Mobile Edge Computing (MEC) [19], and data collection systems [20], [21].

The celebrated merits of NOMA and IRS make them potential candidates for future wireless networks that can enable various IoT applications. Motivated by this, we investigate the potential of integrating IRS and NOMA in uplink communications. Specifically, an optimization based study on the applicability and effectiveness of the introduced system is presented to jointly optimize the users' transmit powers and IRS phase shifts. The aim is to minimize the power consumption of battery-based nodes and to maximize the EE of the system while satisfying the quality-of-service (QoS) and transmit power constraints for each node.

*A. Related Literature on integrated IRS-NOMA*

Much efforts have been devoted into integrating IRS with NOMA. Prior research can be divided into two categories, downlink and uplink IRS-NOMA. Whilst most of the work on this topic is related to downlink IRS-NOMA, there exist only a few articles that consider the uplink scenario. Therefore, in this paper, we are interested in the optimization of the latter scenario by jointly optimizing the users' transmit powers and IRS phase shifts. We first focus on reviewing optimized uplink IRS-NOMA systems, then, for the sake of completeness, we review relevant works on the downlink scenario.

Early work on uplink IRS-NOMA optimization can be found in [22] which considered the users' sum rate maximization under individual QoS and power constraints assuming perfect knowledge of the IRS-BS and IRS-users channels[1]. To achieve their objectives, the authors converted the optimization problem to a semi-definite programming (SDP) problem, and then used SDR to optimize the IRS reflecting coefficients. Besides, the authors in [28] studied the max-min fairness secrecy rate of a two-user IRS assisted uplink NOMA scenario. The solution of the formulated optimization problem includes the conversion of the passive beamforming problem at IRS to SDP followed by employing sequential rank-one constraint relaxation (SROCR) to get a rank-one solution for SDP. In addition, a two-step optimization scheme is devised in [29] for solving the sum rate maximization problem of an active-IRS assisted uplink NOMA scenario. A fixed point iteration (FPI) method has been employed to optimize the phase shifts, which has performance degradation as the number of IRS reflectors increases. In [30], the max-min fairness rate optimization for an IRS assisted uplink NOMA system with receive beamforming at BS was studied. Successive convex approximation (SCA) was used to solve the passive beamforming problem at IRS using second order cone programming (SOCP) in order to avoid the high complexity SDR solution. However, the performance of SCA degrades as the number of IRS reflectors gets high. Moreover, in [31], the EE maximization problem for an IRS assisted uplink multiple-input-multiple-output (MIMO) NOMA scenario was studied. To achieve their research goals, the authors proposed an iterative approach to solve this multi-variable non-convex problem where they used an SDP iterative approach and difference of convex (DC) programming to optimize the IRS phase shifts. Nevertheless, the complexity of their proposed technique is indeed high as it is based on SDP. Amongst the other works in literature, [32] proposed a hybrid NOMA-OMA multiple access scheme in active IRS aided energy-constrained IoT systems to assist the uplink transmission from multiple IoT devices to an access point. The authors' formulated the problem to maximize the sum throughput by optimizing the IRS beamforming vectors across time and resource allocation. In [33], the authors proposed using IRS assisted uplink NOMA transmission to reduce the interference in enhanced mobile broadband (eMBB) and ultra-reliable low-latency communications (URLLC) devices. Both [32], [33] have used the high complexity SDP approach to solve the presented problems. Finally, a low complexity design for IRS assisted uplink sparse code multiple access (SCMA) was proposed in [34], however, the presented design considered SCMA not the general uplink power domain NOMA scheme as our proposed system model in this manuscript.

On the other hand, a range of optimization algorithms were also proposed for optimizing IRS assisted downlink NOMA systems. For instance, the rate performance of IRS assisted downlink NOMA was optimized in [35] by maximizing the minimum decoding signal-to-interference-plus-noise-ratio (SINR) of all the users by leveraging the alternating optimization and SDR techniques to jointly optimize the NOMA power allocation and IRS phase shifts. In addition, in [36], the authors solved the sum rate maximization problem of a downlink multiple-input-single-output (MISO) IRS assisted NOMA system using alternation optimization and SDP-based sequential rank-one relaxation approaches to jointly optimize active beamforming at the BS and passive beamforming at IRS. In [37], the EE of an IRS assisted downlink NOMA system was maximized by jointly optimizing the transmit beamforming at BS and the IRS phase shifts using alternating optimization and SDR. Moreover, the total transmit power minimization at BS for an IRS-empowered downlink MISO NOMA network was investigated in [38], where the joint active beamforming at BS and passive beamforming at IRS problems were formulated as a bi-quadratically constrained optimization problem. Then, the formulated problem was converted to a bi-SDP problem, and a unified DC approach was proposed to get a rank one solution.

It is noteworthy to observe that the formulated optimization problems in most of the previous IRS-NOMA literature have been solved by converting them into SDP problems, and then employing either SDR or successive rank-one procedure to obtain the final solution. However, these solutions suffer from extremely high complexity, which makes them impractical for energy autonomous applications. Moreover, in IRS based systems, the complexity and time requirement for executing existing optimization algorithms severely increase as the number of reflecting elements increases. Therefore, devising low complexity efficient optimization techniques for optimizing IRS assisted uplink NOMA systems is indispensable for energy autonomous networks such as IoT and UAV networks, where BS, possibly attached to a drone, and users have limited hardware capabilities and energy budget.

*B. Motivation and Contributions*

The design of multi user IRS assisted uplink NOMA is challenging since it requires joint optimization of both the transmit powers and the IRS reflection coefficients. The optimization variables are coupled together which makes the problem not tractable for a closed form solution and highly non-convex to be solved using low complexity convex optimization algorithms. Therefore, existing solutions for uplink IRS-NOMA design in the literature use SDP approaches to convert the problem into a convex form to be solved using convex programming tools as mentioned earlier. However, SDP based approaches have high complexity especially when the number of IRS elements are high, which renders these solutions unsuitable for IoT application where the BS may have limited power resources that cannot accommodate very complex tasks.

---

[1]The assumption that the IRS channel vectors is known is widely accepted as there is specific works on IRS channel estimation [23]–[27].

Motivated by the complexity drawback of existing uplink IRS-NOMA optimization algorithms, we propose a low complexity and efficient optimization algorithm that suits battery-powered IoT applications with limited processing capabilities. Specifically, we consider the IRS assisted uplink NOMA system, shown in Fig. 1. Manifold optimization based algorithms are proposed for the system model introduced in this paper aiming at providing a low complexity design with powerful performance. Two main optimization problems, namely, the total users' transmit power minimization and the system's EE maximization, are formulated and solved. It is noteworthy mentioning that these two problems form the basis for reducing power consumption, or increasing energy efficiency in the network. To solve these problems, we use iterative alternation algorithms to jointly optimize the transmit powers of the users and the phase shifts at IRS, under QoS and transmit power constraints for the users. In each iteration of the alternating algorithm, the transmit powers optimization is solved given the IRS passive beamforming coefficients; then, the IRS coefficients are optimized given the obtained users' transmit powers. To solve the IRS phase shifts optimization in each iteration, we devise an efficient complex circle manifold (CCM) optimization algorithm. The performance and complexity of the proposed technique is then compared against three SDP-based benchmarks that were used in the literature of IRS-NOMA, which are SDR [6], SDP-difference of convex (SDP-DC) [38] and SROCR [36]. It is noteworthy to highlight that manifold optimization was used in different IRS assisted communication systems, e.g. [11], [39]–[41]. As for complexity, the proposed CCM technique has a complexity of order $\mathcal{O}(L^2)$, whereas the SDP algorithms have complexity of order $\mathcal{O}(L^{4.5})$, where $L$ is the number of IRS reflectors.

Despite the benefits of integrating NOMA with IRS in energy constrained IoT networks we discussed at the beginning of this subsection, it is necessary to compare the performance gain of integrating IRS-NOMA into energy-constrained IoT systems against the IRS-OMA counterpart. For multi user IRS enabled systems, does the IRS-NOMA scheme always outperform the IRS-OMA one in all cases? To answer this question, we proceed to formulate and address both the challenges of minimizing transmit power and maximizing the EE in uplink OMA system with IRS. This involves the joint optimization of IRS phase shifts and time allocation for each user. The goal is to compare the performance of the IRS-NOMA scheme using the proposed low complexity CCM based algorithm against the optimized IRS-OMA counterpart.

In summary, the main contributions of this work are:

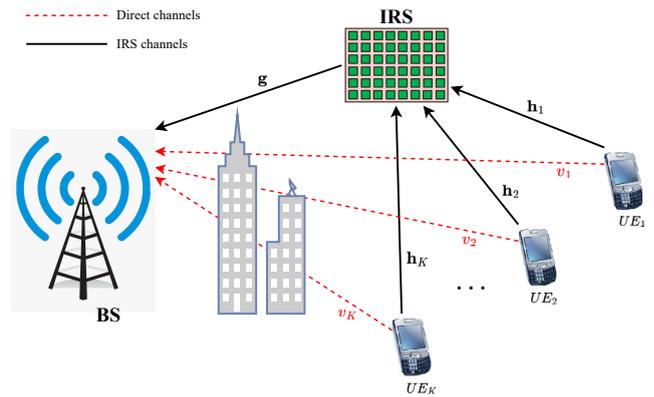

Figure 1: IRS assisted NOMA uplink system model

- The total transmit power minimization and the EE maximization problems are formulated based on the introduced IRS assisted uplink NOMA system. Accordingly, a joint design for optimizing the users' transmit powers and the IRS phase shifts is then proposed to solve each of the two considered problems using alternation optimization. Moreover, we shed light on the relationship between the two considered optimization problems in the simulation results, which has not been thoroughly investigated in the literature. The results unveil that although the two problems converge to the same solution when the users' uplink target rates are high, they have different solutions when the target rates are low.
- Additionally, to solve the IRS passive beamforming sub-problem, we devise a novel low complexity manifold optimization-based approach. The proposed manifold optimization-based algorithm can obtain a locally optimal solution that adheres to the unit-modulus constraints of the IRS reflecting coefficients.
- We further formulate and solve both the transmit power minimization and EE maximization problems for the IRS assisted uplink OMA system, by jointly optimizing the IRS phase shifts and the time allocation for each user, to compare against the IRS assisted NOMA system. Interestingly, the simulation results reveal that NOMA outperforms OMA when the target user rate constraints are high; however, OMA is preferable when the target uplink rate constraints are low.

The rest of the paper is organized as follows. In Sec. II, the IRS-aided uplink NOMA system model is presented, while in Sec. III, the total transmit power minimization problem is formulated and solved. The EE maximization problem is presented and solved in Sec. IV. In Sec. V, we discuss the IRS assisted uplink OMA system. Finally, the results and conclusions are given in Sections VI and VII, respectively.

## II. SYSTEM MODEL

As shown in Fig. 1, we consider an IRS assisted uplink NOMA system in which a group of $K$ single antenna users transmit data symbols to a single antenna BS in the same time-frequency block. The uplink transmission is assisted by an IRS panel which consists of $L$ reflecting elements where the $i$th element can introduce a phase shift $\theta_i$ to the incident signal before reflection. It is worth mentioning that the NOMA scheme is used in this system model to leverage the variation in the users' distances form the BS, i.e., there exist near users and far users, which is known to be an appropriate scenario in which NOMA can perform well. Using NOMA in this setup is beneficial, in most scenarios, since it can provide more power savings than using OMA by reducing the overall transmit power of the users while satisfying their QoS requirements. The dotted lines in Fig. 1 represent the direct channels between the users and the BS. The direct channels are assumed independent and identically distributed (i.i.d.) Rayleigh fading channels because line-of-sight (LoS) paths are not guaranteed in several environments where multiple blockages and obstacles between the users and BS obstruct the communication between them. This actually motivates the deployment of IRS in the considered system model to provide robust links between the users and BS and facilitate communication between them. The direct channel between the $k$th user and BS is denoted as $v_k$. The IRS-BS and the $k$th user-IRS channel vectors are denoted as $\mathbf{g} \in \mathbb{C}^{L \times 1}$ and $\mathbf{h}_k$, respectively, where they are represented by the solid lines in Fig. 1. The IRS panel is typically placed in a position where a LoS path to BS and to the users always exist in practice. Hence, we adopt Rician fading channels to model the channel vectors $\mathbf{g}$ and $\mathbf{h}_k$. The channel vector $\mathbf{g}$ is given as

$$\mathbf{g} = \sqrt{\frac{PL(d_{IB})K_{IB}}{K_{IB}+1}}\mathbf{g}^{LoS} + \sqrt{\frac{PL(d_{IB})}{K_{IB}+1}}\mathbf{g}^{NLoS}, \quad (1)$$

where $K_{IB}$ denotes the Rician factor of $\mathbf{g}$, $d_{IB}$ is the distance between the IRS and the BS, $\mathbf{g}^{LoS}$ and $\mathbf{g}^{NLoS}$ are the line-of-sight (LoS) and non-LoS (NLoS) components, respectively. The LoS component is deterministic, however, the elements of $\mathbf{g}^{NLoS}$ are independent and identically distributed (i.i.d.) complex normal, $\mathcal{CN}(0,1)$, random variables. The channel vector between user $k$ and the IRS, $\mathbf{h}_k$, is given as

$$\mathbf{h}_k = \sqrt{\frac{PL(d_{U_k,I})K_{UI}}{K_{UI}+1}}\mathbf{h}_k^{LoS} + \sqrt{\frac{PL(d_{U_k,I})}{K_{UI}+1}}\mathbf{h}_k^{NLoS}, \quad (2)$$

where $K_{UI}$ denotes the Rician factor of $\mathbf{h}_k$, $d_{U_k,I}$ is the distance between user $k$ and the IRS, $\mathbf{h}_k^{LoS}$ and $\mathbf{h}_k^{NLoS}$ are the LoS and NLoS components, respectively. $PL$ represents the path loss which is modeled for all the channels as

$$PL(d) = \eta_0 \left(\frac{d}{d_0}\right)^{-\alpha}, \quad (3)$$

where $\eta_0$ is the path loss at the reference distance $d_0 = 1$ m, $d$ represents the link distance between the transmitter and the receiver, and $\alpha$ is the path loss exponent.

In Fig. 1 the $K$ users transmit $K$ data symbols *simultaneously*; hence, the received signal at the BS is written as

$$y = \sum_{k=1}^{K}(\mathbf{g}^T\mathbf{W}\mathbf{h}_k + v_k)\sqrt{p_k}s_k + n, \quad (4)$$

where $s_k$ is the data symbol transmitted by user $k$, $\mathbb{E}[|s_k|^2] = 1$, $p_k$ is the transmitted power from user $k$, and $n$ is the complex Gaussian $\mathcal{CN}(0,\sigma_n^2)$ additive white Gaussian noise (AWGN) at the BS. $\mathbf{W}$ is an $L \times L$ diagonal matrix, i.e. $\mathbf{W} = \text{diag}\{\mathbf{w}\}$, whose diagonal elements are the IRS reflection coefficients, where $\mathbf{w} = [e^{j\theta_1}, \ldots, e^{j\theta_L}]^T$. The angle $\theta_i \in [0, 2\pi[$ represents the phase shift introduced by $i$th reflecting element.

In uplink NOMA, the BS performs SIC to successfully decode the $K$ superimposed users' signals. SIC requires ordering of the users according to their effective channel gains. Users with better effective channel gains are decoded first considering the other interfering signals as noise. The combined effective channel of the $k$th user in this system model is $\mathbf{g}^T\mathbf{W}\mathbf{h}_k + v_k$, which clearly depends on the unknown phase shifts matrix $\mathbf{W}$. Therefore, in this paper, we order the users according to their maximum achievable effective channel [35]. For example, for user $k$, we can calculate the maximum achievable effective channel gain at the BS by assuming that the IRS phases are solely optimized to maximize the channel of this user. In this case, the IRS phases should be adjusted to align all the combined IRS channels with the direct link $v_k$ of user $k$, i.e., $\theta_i = \theta_{v_k} - \theta_{g_i} - \theta_{h_{ki}}$. In this case, the maximum achievable channel gain of user $k$ is calculated as $\sum_{i=1}^{L}|g_i||h_{ki}| + |v_k|$. Without loss of generality, we assume that the users are arranged in a descending order as

$$\sum_{i=1}^{L}|g_i||h_{1i}| + |v_1| \geq \ldots \geq \sum_{i=1}^{L}|g_i||h_{Ki}| + |v_K|. \quad (5)$$

Assuming successful decoding of the users at BS using SIC[2], SINR of the $k$th user can be expressed as

$$\gamma_k = \frac{p_k \left|\mathbf{g}^T\mathbf{W}\mathbf{h}_k + v_k\right|^2}{\sum_{j=k+1}^{K} p_j \left|\mathbf{g}^T\mathbf{W}\mathbf{h}_j + v_j\right|^2 + \sigma_n^2}, \quad (6)$$

where $\sum_{j=k+1}^{K} p_j \left|\mathbf{g}^T\mathbf{W}\mathbf{h}_j + v_j\right|^2 = 0$ when $k = K$, i.e., the last user in the SIC order suffers no interference. Then, the corresponding achievable data rate of the $k$th user is given by

$$R_k = \log_2(1+\gamma_k), \quad \forall k. \quad (7)$$

The main interest of this work is to efficiently optimize the IRS reflection coefficients matrix, $\mathbf{W}$, and the users' transmit powers, $p_k$, with the objective of minimize the overall transmit power while satisfying the required data rates. This problem is challenging since the optimization variables, $p_k$ and $\mathbf{W}$, in (6) are coupled, which makes it difficult to solve at once. The optimization problem and its challenges are presented in Sec. III. Existing solutions for similar optimization problems in the literature rely on SDP approaches, which are computationally extensive, especially when the number of IRS elements is large. Motivated by this, we present an efficient manifold optimization-based approach that provides better performance with much lower computational complexity, as illustrated in Sec. III.

---

[2]It is widely accepted to assume the successful decoding of the NOMA users in the capacity equation according to the SIC order [13], [14], [22], [42]–[46]. That is, it is common to assume that decoding can be successful if the user transmits at a rate that does not exceed its capacity limit.

## III. Total Transmit Power Minimization

In this section, we aim to minimize the sum of the transmitted powers of all the users in the multi-user IRS assisted NOMA system assuming that each user has a minimum rate requirement that must be guaranteed to satisfy some QoS requirement at each individual user. This minimization is done by jointly optimizing the passive beamforming coefficients at the IRS, $\mathbf{W}$, and the power allocation at the users. Hence, the sum power minimization problem can be formulated as

$$\min_{\mathbf{p},\mathbf{W}} \quad \sum_{k=1}^{K} p_k \tag{8a}$$
$$\text{s.t.} \quad R_k \geq R_k^{\min}, \qquad k=1,2,\ldots,K, \tag{8b}$$
$$p_k \leq P^{\max}, \tag{8c}$$
$$|w_i|=1, \qquad i=1,2,\ldots,L, \tag{8d}$$

where $w_i = e^{j\theta_i}$ is the $(i,i)$ entry of $\mathbf{W}$ and $P^{\max}$ is the maximum allowable uplink transmit power for the users. Solving the optimization problem in (8) is challenging because the optimization variables, $p_k$ and $w_i$, are multiplied by each other in constraints (8b), and the constraints in (8d) are non-convex. We use the fact that we have two sets of optimization parameters, $p_k$'s and $w_i$'s, coupled together in (8), and propose an alternating optimization algorithm to solve it in $\mathbf{p}$ and $\mathbf{W}$ in an iterative manner.

During each iteration $n$ in our proposed alternating optimization algorithm, we solve the power allocation problem to get $\mathbf{p}^n$ first for a given $\mathbf{W}^{n-1}$, then we optimize $\mathbf{W}$ for a given $\mathbf{p}^n$ to get $\mathbf{W}^n$, and so on. Therefore, the original problem (8) is separated into two sub-problems that are solved alternatively, as explained in the next subsections.

### A. Optimization of users' transmit powers

In the $n$th iteration, given the value of $\mathbf{W}^n$, the power allocation problem in $\mathbf{p}$ reduces to

$$\min_{\mathbf{p}} \quad \sum_{k=1}^{K} p_k \tag{9a}$$
$$\text{s.t.} \quad C_k(\mathbf{W}^n,\mathbf{p}) = p_k \left|\mathbf{g}^T \mathbf{W}^n \mathbf{h}_k + v_k\right|^2 - (2^{R_k^{\min}} - 1)$$
$$\times \left( \sum_{j=k+1}^{K} p_j \left|\mathbf{g}^T \mathbf{W}^n \mathbf{h}_j + v_j\right|^2 + \sigma_n^2 \right) \geq 0, \tag{9b}$$
$$k=1,2,\ldots,K,$$
$$p_k \leq P^{\max}, \tag{9c}$$
$$p_k \geq 0. \tag{9d}$$

It is worth noting that there is no need for adding extra constraint for the SIC decoding order since it is already embedded in the rate constraints of the NOMA users (9b). That is, given the reflection coefficient matrix $\mathbf{W}$, the constraints in (9b) must be satisfied with equality as an optimality condition, which guarantees the correct SIC order.

The optimization problem (9) can be expressed as a linear program in $\mathbf{p}$ in the form of [47]

$$\min_{\mathbf{x}} \quad \mathbf{c}^T \mathbf{x} \tag{10a}$$
$$\text{s.t.} \quad \mathbf{A}\mathbf{x} \leq \mathbf{b} \tag{10b}$$
$$\mathbf{x} \geq \mathbf{0}, \tag{10c}$$

where the constraints in (9b) and (9c) are affine inequalities in $\mathbf{p}$ which can be all written in the form of (10b), and the cost function (9a) is linear having the form of (10a) with $\mathbf{c}$ equals to an all ones vector. Therefore, problem (9) is a linear program in $\mathbf{p}$ which can be easily solved using the standard *simplex method*. The simplex method is a well known algorithm to solve any linear program, having a linear objective function and affine constraints, using slack variables and pivot variables as a means to find the optimal solution [47], [48].

In the next subsection, we present a novel and low complexity approach for solving the challenging optimization problem (8) in $\mathbf{W}$ given the values of $p_k$'s. We first present an SDP approach as a benchmark from the current solutions in the literature, followed by our proposed novel and low complexity CCM based optimization approach. The performance of the two approaches will be compared along with other current SDP schemes, in the results section.

### B. Optimization of the IRS reflection coefficients

Firstly, we present a current SDR benchmark scheme and emphasize the drawbacks of existing counterparts.

For some given $p_k$ values in the $n$th iteration, the optimization problem in (8) becomes a feasibility-check problem in $\mathbf{W}$, since the cost function in (8a) does not depend on the IRS coefficients $\mathbf{W}$. Hence, given the power allocation vector $\mathbf{p}^n$, the feasibility-check problem in $\mathbf{W}$ is given as

$$\text{Find} \quad \mathbf{W} \tag{11a}$$
$$\text{s.t.} \quad C_k(\mathbf{W},\mathbf{p}^n) \geq 0, \qquad k=1,2,\ldots,K, \tag{11b}$$
$$|w_i|=1, \qquad i=1,2,\ldots,L. \tag{11c}$$

*1) SDR benchmark for optimizing $\mathbf{W}$:* Optimizing $\mathbf{W}$ in problem (11) can be solved using SDR by converting problem (11) into a SDP as follows. Problem (11) is non-convex because of the non-convex constraints in (11b) and the unit modulus constraints in (11c). However, by converting (11) into a SDP, the problem is transformed to a convex problem and solved using standard tools such as CVX. The constraints in (11b) can be converted into quadratic forms w.r.t. the passive beamforming vector $\mathbf{w}$, then converting them to convex affine constraints using the SDP technique as follows. The squared magnitude of the effective channel of user $k$ can be written as

$$\left|\mathbf{g}^T \mathbf{W} \mathbf{h}_k + v_k\right|^2 = \left|\mathbf{g}^T \mathbf{H}_k \mathbf{w} + v_k\right|^2, \tag{12}$$

where $\mathbf{H}_k = \text{diag}(\mathbf{h}_k)$ and $\mathbf{w} \in \mathbb{C}^{L \times 1}$ is a column vector containing the diagonal elements of $\mathbf{W}$. Then, we can expand the squared magnitude of the effective channel of the $k$th user as

$$\left|\mathbf{g}^T \mathbf{H}_k \mathbf{w} + v_k\right|^2 = (\mathbf{w}^H \mathbf{H}_k^H \mathbf{g}^* + v_k^*)(\mathbf{g}^T \mathbf{H}_k \mathbf{w} + v_k) \tag{13}$$
$$= (\mathbf{w}^H \mathbf{z}_k + v_k^*)(\mathbf{z}_k^H \mathbf{w} + v_k), \tag{14}$$

where $\mathbf{z}_k = \mathbf{H}_k^H \mathbf{g}^* \in \mathbb{C}^{L \times 1}$. Hence, the squared effective channel gain of user $k$ can be given as a quadratic term in $\mathbf{w}$ as

$$\left|\mathbf{g}^T \mathbf{H}_k \mathbf{w} + v_k\right|^2 = \mathbf{w}^H \mathbf{z}_k \mathbf{z}_k^H \mathbf{w} + \mathbf{w}^H \mathbf{z}_k v_k + \mathbf{z}_k^H \mathbf{w} v_k^* + |v_k|^2. \quad (15)$$

Equation (15) can be expressed as $\bar{\mathbf{w}}^H \mathbf{B}_k \bar{\mathbf{w}} + |v_k|^2$, where

$$\bar{\mathbf{w}} = \begin{bmatrix} \mathbf{w} \\ 1 \end{bmatrix} \text{ and } \mathbf{B}_k = \begin{bmatrix} \mathbf{z}_k \mathbf{z}_k^H & \mathbf{z}_k v_k \\ \mathbf{z}_k^H v_k^* & 0 \end{bmatrix}. \quad (16)$$

Knowing that $\bar{\mathbf{w}}^H \mathbf{B}_k \bar{\mathbf{w}} = \text{trace}(\bar{\mathbf{w}}^H \mathbf{B}_k \bar{\mathbf{w}}) = \text{trace}(\mathbf{B}_k \bar{\mathbf{w}} \bar{\mathbf{w}}^H)$, and by letting $\widehat{\mathbf{W}} = \bar{\mathbf{w}} \bar{\mathbf{w}}^H \in \mathbb{C}^{L+1 \times L+1}$, we can convert the problem into SDP. The matrix $\widehat{\mathbf{W}}$ is the outer product of the vector $\bar{\mathbf{w}}$ with itself, hence $\widehat{\mathbf{W}}$ is a symmetric positive semi-definite matrix whose rank is 1. Consequently, by introducing the slack variable $\alpha$ to maximize the minimum of the constraints in (11b), problem (11) can be reformulated as

$$\max_{\widehat{\mathbf{W}}, \alpha} \quad \alpha \quad (17a)$$

$$\text{s.t.} \quad p_k \left( \text{trace}\{\mathbf{B}_k \widehat{\mathbf{W}}\} + |v_k|^2 \right) - (2^{R_k^{\min}} - 1) \times$$

$$\left( \sum_{j=k+1}^{K} p_j \left( \text{trace}\{\mathbf{B}_j \widehat{\mathbf{W}}\} + |v_j|^2 \right) + \sigma_n^2 \right) - \alpha \geq 0, \forall k, \quad (17b)$$

$$\widehat{\mathbf{W}}(i,i) = 1, \quad i = 1, 2, \ldots, L+1, \quad (17c)$$

$$\widehat{\mathbf{W}} \succeq \mathbf{0}, \quad \text{rank}(\widehat{\mathbf{W}}) = 1, \quad (17d)$$

$$\alpha \geq 0. \quad (17e)$$

It should be noted that introducing the slack variable $\alpha$ in (17) is a standard method for maximizing the minimum of some functions [6]. Obviously, problem (17) is a SDP, however, still non-convex because of the non-convex rank one constraint in (17d). By dropping (17d), SDP becomes convex and can be solved using CVX. Assuming the optimal solution of the SDR of (17) is $\widehat{\mathbf{W}}^*$, then we need to convert $\widehat{\mathbf{W}}^*$ to a feasible solution of the original problem (11). The rank of the obtained solution, $\widehat{\mathbf{W}}^*$, of the SDR is generally greater than 1, thus we will always need to reduce our solution to a rank one matrix. An effective way to find a good rank one approximation is to select the eigenvector of $\widehat{\mathbf{W}}^*$ that corresponds to its maximum eigenvalue [49]. Assuming that the maximum eigenvalue of $\widehat{\mathbf{W}}^*$ is $\lambda_1$ and the corresponding eigenvector is $\mathbf{q}_1$, then $\widehat{\mathbf{w}} = \sqrt{\lambda_1} \mathbf{q}_1$ can be considered a solution to our original problem (11). However, this solution maybe still infeasible and does not satisfy the unit modulus constraints for $w_i$'s. To map the solution to a nearby feasible solution, the elements of $\widehat{\mathbf{w}}$ are normalized to $\widetilde{w}_i = \frac{\widehat{w}_i}{|\widehat{w}_i|}, \forall i \in \{1, \ldots, L\}$.

We can see that the SDP approaches broaden the original feasibility region to a larger set. This expansion is achieved by transforming the optimization variable from a vector, $\mathbf{w} \in \mathbb{C}^{L \times 1}$, to an $L \times L$ matrix, denoted as $\mathbf{W} = \mathbf{w}\mathbf{w}^H$, which is the outer product of the original optimization vector. Subsequently, the optimal solution of the newly expanded SDP problem, denoted as $\mathbf{W}^*$, undergoes projection onto the original feasible set to derive a solution for the original optimization vector, $\mathbf{w}^*$. The projection is executed through Cholesky decomposition, which may, sometimes, yield results deviating from the optimal solution to the original problem. This deviation contributes to a relative reduction in performance when compared to the proposed CCM-based approach.

Moreover, the complexity of existing convex SDP based solutions are extremely high since they search over a far larger feasible set with $L^2$ optimization variables which are the elements of the optimization matrix, $\mathbf{W}$. However, as illustrated next, the proposed CCM based algorithm searches over the original feasible set which is the complex circle manifold with far less number of optimization variables, i.e., $L$, which are the elements of the original optimization vector, $\mathbf{w}$. This leads to a substantial complexity reduction in our algorithm compared to the counterparts as shown in Sec. III-D.

*2) The proposed CCM based approach for optimizing $\mathbf{W}$:* In this subsection, we introduce an efficient and low complexity solution for optimizing the passive beamforming vector in (11), which is the proposed manifold optimization based algorithm. As long as the cost function in (8a) does not depend on $\mathbf{W}$, we propose to solve (11) in the second step of each iteration of the proposed alternating optimization algorithm to expand the feasible region of (9) in the first step of the next iteration, i.e., when optimizing $\mathbf{p}$ given the obtained $\mathbf{W}$. To enhance the convergence of finding a solution to (11), the problem can be further transformed into an optimization problem with a clear objective function to find the IRS phase shifts, $\mathbf{W}$, which further reduces the total uplink transmit power. The logic of this is that all the constraints are already satisfied with equality by the optimal solution of the sum power minimization sub-problem in (9). Therefore, optimizing $\mathbf{W}$, to enforce the users' achievable data rates to be larger than the minimum target rates forthrightly results in the reduction of the total transmit power in (9). More specifically, we need to find a solution for $\mathbf{W}$ that maximizes the minimum of the constraints in (11b). Therefore, the problem of maximizing the minimum of the constraints shall be written as

$$\max_{\mathbf{W}} \quad \min\{C_1(\mathbf{W}, \mathbf{p}^n), C_2(\mathbf{W}, \mathbf{p}^n), \ldots, C_K(\mathbf{W}, \mathbf{p}^n)\} \quad (18)$$

$$\text{s.t.} \quad (11b), (11c).$$

The $\min$ function in (18) is not smooth. Hence, we use a smooth and differentiable approximation as

$$\min_{k} |C_k| = \left( \sum_{k=1}^{K} C_k^{-\omega} \right)^{-1/\omega}, \text{ as } \omega \to +\infty, \quad (19)$$

and $|C_k| = C_k$ because we know that all the constraints are non negative, i.e., $C_k \geq 0$. Therefore, the manifold optimization problem can be written as

$$\max_{\mathbf{W}} \quad \left( \sum_{k=1}^{K} C_k^{-\omega} \right)^{-1/\omega} \quad (20)$$

$$\text{s.t.} \quad (11b), (11c).$$

The unit-modulus constraints in (11c) restrict the optimization vector, $\mathbf{w}$, to be located on the surface of a smooth Riemannian manifold contained in $\mathbb{C}^L$. Precisely, all the optimization variables, $w_i$, lie on a continuous surface called the CCM,

which is defined as
$$\mathcal{S} = \{w_i \in \mathbb{C} : |w_i| = 1\}. \quad (21)$$

The circle, $\mathcal{S}$, forms a smooth sub-manifold of $\mathbb{C}$ which is a Riemannian manifold. Due to having $L$ optimization variables in (20), the feasible set of the optimization problem is the Cartesian product of $L$ complex circles which is given as
$$\mathcal{S}_1 \times \mathcal{S}_2 \times \ldots \times \mathcal{S}_L. \quad (22)$$

The Cartesian product of smooth Riemannian manifolds forms a smooth Riemannian sub-manifold of $\mathbb{C}^L$. Hence, the feasible set of (20) is an $L$-dimensional CCM which is formally defined as
$$\mathcal{S}^L \triangleq \mathcal{S}_1 \times \ldots \times \mathcal{S}_L = \{\mathbf{w} = [w_1, \cdots, w_L] \in \mathbb{C}^L \\ : |w_1| = \cdots = |w_L| = 1\}. \quad (23)$$

The optimization problem in (20) has extra constraints in (11b) other than the unit-modulus constraints. In the following, we explain how to mnanage the constraints given in (11b), to solve (20). We propose to handle these additional constraints by using a standard approach called the exact penalty method. To account for the constraints in (11b), the exact penalty method adds a weighted penalty term for each constraint to the the objective function being optimized. When one of the constraints is violated, the modified objective function is largely penalized by moving far away from the optimum point. By using the exact penalty method, (20) is reduced to an unconstrained optimization problem, but generally having a non-smooth objective function. Therefore, in the Riemannian case, the exact penalty method transforms (20) to an unconstrained optimization problem over the $L$-dimensional complex circle manifold as follows
$$\max_{\mathbf{w} \in \mathcal{M}} \left(\sum_{k=1}^{K} C_k(\mathbf{w})^{-\omega}\right)^{-1/\omega} + \rho \left(\sum_{k=1}^{K} \max\{0, -C_k(\mathbf{w})\}\right), \quad (24)$$
where $\rho > 0$ is a penalty weight and $\mathcal{M}$ is the $L$-dimensional complex circle manifold over which the problem is optimized. Note that the constant modulus constraints in (11c) are satisfied by restricting the feasible set to the manifold $\mathcal{M}$. In the exact penalty method, only a finite penalty weight, $\rho$, is needed to exactly satisfy the constraints, hence the method's name, [50]. The penalized objective function in (24) is not smooth due to the existence of the max functions that replace the constraints with the corresponding penalties. Therefore, we can use a smoothing technique to smooth and solve (24). By using the linear-quadratic loss approach [51], the max function in (24) can be approximated, using a smoothing parameter $u > 0$, as $\max\{0, x\} \approx \mathcal{P}(x, u)$, where $\mathcal{P}(x, u)$ is given by
$$\mathcal{P}(x, u) = \begin{cases} 0 & x \leq 0 \\ \frac{x^2}{2u} & 0 \leq x \leq u \\ x - \frac{u}{2} & x \geq u, \end{cases} \quad (25)$$

Hence, a smooth unconstrained version of our manifold optimization problem can be written as
$$\max_{\mathbf{w} \in \mathcal{M}} \quad Q(\mathbf{w}) = \left(\sum_{k=1}^{K} C_k(\mathbf{w})^{-\omega}\right)^{-1/\omega} + \rho \left(\sum_{k=1}^{K} \mathcal{P}(-C_k(\mathbf{w}), u)\right). \quad (26)$$

Thus, the problem in (20) is converted to a smooth unconstrained manifold optimization problem whose feasible points lie on the surface of the complex circle manifold, $\mathcal{S}^L$. In that case, we can utilize gradient-based manifold optimization techniques to solve (26).

Similar to the case of Euclidean spaces, there are two main steps for a gradient-descent based algorithm on Riemannian manifolds. Firstly, a descent direction should be found, then a step size is computed along this direction. Afterwards, the solution is updated iteratively until convergence by repeating the two steps in each iteration. However, calculating a descent direction should be adjusted to take into account the geometric nature of the manifold, which is discussed in the following. The tangent space, $T_{\mathbf{w}}\mathcal{M}$, at a point, $\mathbf{w}$, on a differentiable manifold, $\mathcal{M}$, is defined as the real vector space that intuitively contains the possible directions in which one can tangentially pass through $\mathbf{w}$. The tangent space at $\mathbf{w}$ is given by
$$T_{\mathbf{w}}\mathcal{M} = \{\mathbf{v} \in \mathbb{C}^L : \Re(\mathbf{v} \odot \mathbf{w}^*) = \mathbf{0}_L\}, \quad (27)$$
where $\Re(.)$ denotes the element-wise real-part of the complex vector, and $\odot$ is the Hadamard element-wise multiplication. In the context of manifold optimization, the gradient of a function is called the *Riemannian gradient* which is defined as the direction of the steepest increase of the objective function at a certain point $\mathbf{w}$, and this direction must lie in the tangent space at the point $\mathbf{w}$. The Riemannian gradient at a point on the manifold is calculated by first calculating the Euclidean gradient at that point then projecting it onto the tangent space at the point using a projection operator. The orthogonal projection operator of a vector $\mathbf{v}$ onto the tangent space, $T_{\mathbf{w}}\mathcal{M}$, at point $\mathbf{w}$ on the $L$-dimensional complex circle manifold is given by [50]
$$P_{T_{\mathbf{w}}\mathcal{M}}(\mathbf{v}) = \mathbf{v} - \Re(\mathbf{v} \odot \mathbf{w}^*) \odot \mathbf{w}. \quad (28)$$

Hence, the Riemannian gradient of our smooth objective function $Q$ in (26) on the manifold can be given as
$$\nabla_{\mathcal{M}} Q(\mathbf{w}) = P_{T_{\mathbf{w}}\mathcal{M}}(\nabla Q(\mathbf{w})) \\ = \nabla Q(\mathbf{w}) - \Re(\nabla Q(\mathbf{w}) \odot \mathbf{w}^*) \odot \mathbf{w}, \quad (29)$$
where $\nabla Q(\mathbf{w})$ is the Euclidean gradient at the point $\mathbf{w}$. In the following, we derive the the Euclidean gradient of $Q$, which is required to obtain the Riemannian gradient. The Euclidean gradient is given by
$$\nabla Q(\mathbf{w}) = \left[\frac{\partial Q}{\partial w_1} \frac{\partial Q}{\partial w_2} \cdots \frac{\partial Q}{\partial w_L}\right]^T, \quad (30)$$
where $\partial Q / \partial w_i = \partial Q / \partial \Re(w_i) + j \partial Q / \partial \Im(w_i)$, and $\Im(.)$ denotes the imaginary part of the complex number. Every partial derivative w.r.t. $w_i$ is calculated as
$$\frac{\partial Q}{\partial w_i} = \frac{\partial f}{\partial w_i} + \rho \sum_{k=1}^{K} \frac{\partial}{\partial w_i} \mathcal{P}(-C_k(\mathbf{w}), u), \quad (31)$$

where $f$ is the first term of the objective function $Q$ in (26), and $\frac{\partial f}{\partial w_i}$ is computed as

$$\frac{\partial f}{\partial w_i} = \frac{\partial f}{\partial \mathfrak{R}(w_i)} + j\frac{\partial f}{\partial \mathfrak{I}(w_i)}$$
$$= \left(\sum_{k=1}^{K} C_k(\mathbf{w})^{-\gamma}\right)^{-1/\gamma-1} \sum_{k=1}^{K} C_k(\mathbf{w})^{-\gamma-1} C'_{ki}(\mathbf{w}), \quad (32)$$

where $C'_{ki}(\mathbf{w})$ is calculated as

$$C'_{ki}(\mathbf{w}) = \frac{\partial C_k(\mathbf{w})}{\partial \mathfrak{R}(w_i)} + j\frac{\partial C_k(\mathbf{w})}{\partial \mathfrak{I}(w_i)}$$
$$= p_k\Big(2\mathfrak{R}(\mathbf{d}_k^T\mathbf{w})(a_{ki}-jb_{ki}) + 2\mathfrak{I}(\mathbf{d}_k^T\mathbf{w})(b_{ki}+ja_{ki})\Big)$$
$$- (2^{R_k^{\min}}-1)\Bigg\{\sum_{l=k+1}^{K} p_l\Big(2\mathfrak{R}(\mathbf{d}_k^T\mathbf{w})(a_{li}-jb_{li})$$
$$+ 2\mathfrak{I}(\mathbf{d}_k^T\mathbf{w})(b_{li}+ja_{li})\Big)\Bigg\}, \quad (33)$$

where $a_{ki}=\mathfrak{R}(h_{ki}g_i)$, $b_{ki}=\mathfrak{I}(h_{ki}g_i)$ and the vector $\mathbf{d}_k = \mathbf{h}_k \odot \mathbf{g}$ is the Hadamard product of the two vectors $\mathbf{h}_k$ and $\mathbf{g}$. The partial derivative, $\frac{\partial \mathcal{P}(-C_k(\mathbf{w}),u)}{\partial w_i}$, is calculated as

$$\frac{\partial \mathcal{P}(-C_k(\mathbf{w}),u)}{\partial w_i} = \begin{cases} 0 & -C_k(\mathbf{w},u) \leq 0 \\ \frac{C_k(\mathbf{w})}{u}C'_{ki}(\mathbf{w}) & 0 \leq -C_k(\mathbf{w},u) \leq u \\ -C'_{ki}(\mathbf{w}) & -C_k(\mathbf{w},u) \geq u, \end{cases}$$
(34)

where $C'_{ki}(\mathbf{w})$ is calculated in (33). Then, by substituting (32) and (34) in (31), we obtain the Euclidean gradient which is required to calculate the Riemannian gradient in (29) for our algorithm. **Algorithm 1** illustrates the steps of solving the unconstrained manifold optimization problem in (26) by updating the penalty coefficient $\rho$ and the smoothing parameter $u$ in an iterative manner.

The feasibility of (20) is ensured by restricting the feasible set of the optimization problem to the complex circle manifold to satisfy the unit modulus constraints in (11c). The other constraints in (11b) are satisfied by penalizing the original cost function as in (24) using the exact penalty method.

Next, we investigate the behavior of **Algorithm 1** and explain how the parameters $\rho$ and $u$ are updated in each iteration until convergence. First, we need to mention that the optimum points of the penalized unconstrained problem in (26) are the same as the optimum points of the original problem in (20) when the penalty coefficient $\rho$ is higher than a certain threshold [50]. This threshold is usually unknown; hence, we adopt a common iterative approach provided in [52] to address this problem. This approach does not start with a large $\rho$ because this may slow down the convergence of **Algorithm 1**. Therefore, **Algorithm 1** starts with a relatively low initial value of $\rho_0$, then this value is increased in each iteration, by multiplying it by the constant $\theta_\rho > 1$, if the constraints in (11b) are violated as shown in lines 8 and 9 in **Algorithm 1**. The parameter $\tau$ is a tolerance factor which is a low positive number so that if $-C_k(\mathbf{w}_{l+1})$ surpasses $\tau$, the point $\mathbf{w}_{l+1}$ is considered out of the feasible region and $\rho$ must be increased.

The approximation function in (34) is more accurate when

---

**Algorithm 1:** Exact penalty method via smoothing

1 **Input:** Starting point $\mathbf{w}_0$, starting penalty coefficient $\rho_0$, starting smoothing accuracy $u_0$, minimum smoothing accuracy $u_{\min}$, constants $\theta_u \in (0,1)$, $\theta_\rho > 1$, $\tau \geq 0$, minimum step length $d_{\min}$.
2 **for** $l = 0, 1, 2, \ldots$ **do**
3    To obtain $\mathbf{w}_{l+1}$, choose any sub-solver to approximately solve
$$\min_{\mathbf{w} \in \mathcal{M}} \quad Q(\mathbf{w}, \rho_l, u_l)$$
with warm-start at $\mathbf{w}_l$ and stopping criterion
$$\|\text{grad } Q(\mathbf{w}, \rho_l, u_l)\| \leq \delta.$$
4    **if** $(\text{dist}(\mathbf{w}_l, \mathbf{w}_{l+1}) < d_{\min}$ or $u_l \leq u_{\min})$ and $C_k(\mathbf{w}_{l+1}) < \tau$ **then**
5       Return $\mathbf{w}_{l+1}$;
6    **end**
7    $u_{l+1} = \max\{u_{\min}, \theta_u u_l\}$;
8    **if** $(l = 0$ or $-C_k(\mathbf{w}_{l+1}) \geq \tau)$ **then**
9       $\rho_{l+1} = \theta_\rho \rho_l$
10    **else**
11       $\rho_{l+1} = \rho_l$;
12    **end**
13 **end**

---

the smoothing parameter $u_l$ is decreased. However, if $u_l$ is too small, numerical difficulties may arise when using the approximation function in (34). Consequently, **Algorithm 1** starts with $u_0$, then keeps decreasing $u_l$, by multiplying it with the constant fraction $\theta_u$, in each iteration, as in line 7, until it reaches a minimum value $u_{\min}$. **Algorithm 1** terminates when the Euclidean distance between $\mathbf{w}_{l+1}$ and $\mathbf{w}_l$ is smaller than $d_{\min}$. To solve the unconstrained optimization problem in line 3 in each iteration, we use the trust region manifold optimization solver [50], that exists in *Manopt* MATLAB toolbox [53], by setting a stopping criterion on the gradient norm.

### C. Convergence of the alternating optimization discussion

The complete procedure of the proposed CCM based alternation optimization algorithm is summarized in **Algorithm 2** where we alternately solve (8) by solving the power allocation problem (9) and the passive beamforming problem at IRS (26) using the proposed CCM approach. In each iteration of **Algorithm 2**, we get a solution that is used as the initial point for the next iteration. Specifically, the alternating optimization algorithm starts by solving problem (9) given the beamforming vector, $\mathbf{w}_{n-1}$, rather than solving (26) given $\mathbf{p}_{n-1}$. This is intentionally designed because (9) is always feasible in $\mathbf{p}$ for any beamforming vector $\mathbf{w}$, however, (26) may not always be feasible for any arbitrary $\mathbf{p}$. Intuitively, when the obtained solution from (26) achieves a higher user uplink rate than the target minimum rate constraint for user $k$, then the transmit power of user $k$ can be suitably decreased while maintaining all the rate constraints. Consequently, the total transmit power in (9) can be reduced accordingly.

More precisely, we can guarantee the convergence of the proposed alternating optimization solution for (8) using the following argument. We can deduce that the cost function of (9), which is the sum transmit powers at the users, always decreases over the iterations when solving (9) and (26) iteratively using **Algorithm 2**. This proposition can be proven as follows. When solving (26) in the $n$-th iteration given the transmit power vector $\mathbf{p}_n$, then the pair $(\mathbf{p}_n, \mathbf{w}_n)$ must be feasible and satisfy the constraints of the original problem (8). This is simply because both the sub-problems (9) and (26) have the same constraints of (8). However, the obtained $\mathbf{w}_n$ expands the feasibility region of (9) in the $(n+1)$-th iteration of **Algorithm 2**, which directly leads to a solution of $\mathbf{p}_{n+1}$ that corresponds to a lower cost function of (9), and so on. The algorithm keeps decreasing the cost function of (9) over the iterations until, during some iteration, problem (26) cannot expand the feasibility region anymore by increasing the achievable SINR of the users. Therefore, (26) will hold on giving the same solution for $\mathbf{w}$ as the previous iteration, which causes **Algorithm 2** to converge at such iteration with no further reduction in the cost function of (9).

In other words, Assume that the objective function of (9), based on the a certain IRS coefficients vector $\mathbf{w}$, is denoted as $f(\mathbf{w}, \mathbf{p})$. In the $n$-th iteration of **Algorithm 2**, if a feasible solution to (26) exists, i.e. $(\mathbf{w}^n, \mathbf{p}^n)$, then this solution is also feasible to problem (9). Then, it follows that $f(\mathbf{w}^n, \mathbf{p}^{n+1}) \overset{(a)}{\leq} f(\mathbf{w}^n, \mathbf{p}^n) \overset{(b)}{=} f(\mathbf{w}^{n-1}, \mathbf{p}^n)$; $(a)$ follows because for a given $\mathbf{w}^n$, $\mathbf{p}^{n+1}$ is the optimal solution to problem (9). Moreover, $(b)$ follows since the objective function of (9) does not depend on $\mathbf{w}$ and is only affected by $\mathbf{p}$.

---

**Algorithm 2:** Sum power minimization alternation optimization algorithm

1 **Initialize iteration number:** $n = 0$.
2 **Initial feasible point:** $\mathbf{p}_0, \mathbf{W}_0$.
3 **while** $\Xi \geq \epsilon$ **do**
4     Solve (9) in $\mathbf{p}$ given fixed IRS coefficients matrix $\mathbf{W}_{n-1}$;
5     Solve (26) in $\mathbf{W}$, using **Algorithm 1**, given fixed users' power allocation vector $\mathbf{p}_n$;
6     Calculate $\Xi = \sum_{k=1}^{K} p_{k,n-1} - \sum_{k=1}^{K} p_{k,n}$;
7     $n \to n+1$;
8 **end**
9 **Return:** $\mathbf{p}^*, \mathbf{W}^*$.

---

*D. Complexity analysis*

In each iteration of the alternation optimization algorithm, we solve the passive beamforming optimization at the IRS either with the proposed manifold optimization-based method in **Algorithm 1** or with the SDR benchmark. The complexity of the SDR sub-problem in (17) is determined by the complexity of the interior point method used for solving the semi definite program, which is given as $\mathcal{O}(\max(L+1, K)^4 (L+1)^{1/2} \log(1/\zeta))$ [54], where $\zeta$ is the accuracy of the solution.

Now, we then analyze the complexity of the proposed manifold optimization-based method presented in **Algorithm 1**. The complexity of **Algorithm 1** is determined by the complexity of solving the unconstrained manifold optimization problem in line 3 in each iteration. The complexity of the solving the manifold optimization problem mainly depends on calculating the Euclidean gradient of the cost function which is given as $\mathcal{O}(L^2)$ [39], [40]. As long as our optimization problem has $K$ constraints for the users, the complexity of calculating the Euclidean gradient of $Q$ in (30) can be given as $\mathcal{O}(KL^2)$. Therefore, the overall complexity of **Algorithm 1** can be written as $\mathcal{O}(\Omega_{Alg1} K L^2)$, where $\Omega_{Alg1}$ is the number of iterations of **Algorithm 1**.

From the above discussion, we can easily deduce that the proposed manifold optimization-based algorithm has much lower complexity, i.e., order 2, with $L$ compared to the SDR benchmark whose complexity is of order 4.5.

## IV. ENERGY EFFICIENCY MAXIMIZATION

In this section, EE maximization is discussed for the uplink NOMA scenario in Fig. 1. The EE of our considered system model is defined as the ratio of the achievable sum-rate of the $K$ users in the NOMA resource block over the total sum of the consumed transmission power by the users. Our objective in this section is to maximize EE while guaranteeing the minimum QoS constraint at each user, i.e., $R_k \geq R_k^{\min}$. The achievable sum rate of the NOMA scheme can be rewritten as

$$R^{sum} = \sum_{k=1}^{K} \log_2 \left( 1 + \frac{p_k \left| \mathbf{g}^T \mathbf{W} \mathbf{h}_k + v_k \right|^2}{\sum_{j=k+1}^{K} p_j \left| \mathbf{g}^T \mathbf{W} \mathbf{h}_j + v_j \right|^2 + \sigma_n^2} \right)$$

$$= \log_2 \left( 1 + \frac{\sum_{k=1}^{K} p_k \left| \mathbf{g}^T \mathbf{W} \mathbf{h}_k + v_k \right|^2}{\sigma_n^2} \right). \quad (35)$$

In this manuscript, we adopt the practical power consumption model used in [7] to model the total power dissipated in our IRS-NOMA system. The considered total power consumption model can be expressed as [7]

$$P_{\text{total}} = \sum_{k=1}^{K} (\nu_k^{-1} p_k + P_{\text{UE},k}) + P_{\text{BS}} + L P_l, \quad (36)$$

where $\nu_k$ is the efficiency of the transmit power amplifiers of the users' devices, $P_{\text{UE},k}$ is the hardware static power consumption at the users, $P_{\text{BS}}$ is the hardware static power dissipation at BS, and $P_l$ denotes the power consumption of each reflection element of the IRS. Therefore, our considered EE maximization is formulated as

$$\max_{\mathbf{p}, \mathbf{w}} \quad \frac{\log_2 \left( 1 + \frac{\sum_{k=1}^{K} p_k \left| \mathbf{g}^T \mathbf{H}_k \mathbf{w} + v_k \right|^2}{\sigma_n^2} \right)}{P_{\text{total}}} \quad (37a)$$

$$\text{s.t.} \quad p_k \leq P^{\max}, \quad (37b)$$

$$C_k(\mathbf{w}, \mathbf{p}) \geq 0, \quad k = 1, 2, \ldots, K, \quad (37c)$$

$$|w_i| = 1, \quad i = 1, 2, \ldots, L, \quad (37d)$$

where $\mathbf{H}_k = \text{diag}(\mathbf{h}_k)$, $P^{\max}$ is the maximum transmit power allowed at the users, and $\mathbf{w}$ is a column vector that contains the diagonal elements of $\mathbf{W}$. We propose to solve (37) using alternating optimization as in the previous section because we have two blocks of optimization parameters $\mathbf{p}$ and $\mathbf{w}$. In the next subsections, we first solve the problem in $\mathbf{p}$ given $\mathbf{w}$, then the problem is solved in $\mathbf{w}$ given $\mathbf{p}$.

### A. Optimizing the transmit power vector

Given the beamforming vector $\mathbf{w}$, the optimizing (37) is a non-linear fractional program in $\mathbf{p}$ which can be expressed as

$$\max_{\mathbf{p}} \quad \frac{\log_2\left(1 + \frac{\sum_{k=1}^{K} p_k a_k}{\sigma_n^2}\right)}{P_{\text{total}}} \tag{38a}$$

$$\text{s.t.} \quad p_k \leq P^{\max}, \tag{38b}$$

$$C_k(\mathbf{p}; \mathbf{w}) \geq 0, \quad k = 1, 2, \ldots, K, \tag{38c}$$

where $a_k = |\mathbf{g}^T \mathbf{H}_k \mathbf{w} + v_k|^2$. The feasible set of (38) is convex since all the constraints are affine in $\mathbf{p}$. The fractional objective function has a concave numerator w.r.t. $p_k, k \in \{1, \ldots, K\}$, while the denominator is an affine mapping over the feasible set of $\mathbf{p}$. Therefore, this non-linear fractional program is *pseudo-concave* and has an optimal solution that can be efficiently obtained using the iterative Dinkelbach's algorithm [55]. The specific procedure of this algorithm is summarized in **Algorithm 3**. During each iteration, the proposed algorithm needs to solve the following problem

$$\max_{\mathbf{p}} \quad \log_2\left(1 + \frac{\sum_{k=1}^{K} p_k a_k}{\sigma_n^2}\right) - \beta (P_{\text{total}}) \tag{39}$$

$$\text{s.t.} \quad (38b), (38c),$$

where $\beta$ is a known constant. Obviously, (39) is concave and its feasible set is convex, hence it can be solved using standard optimization algorithms like the interior point method. Nonetheless, (39) has a specific structure that can be exploited to develop a low complexity solution. The standard approaches are computationally intensive especially that solving (39) is repeated in each iteration of **Algorithm 3**.

---

**Algorithm 3:** EE max. for $\mathbf{p}$

1 **Initialize parameters:** $\epsilon > 0$, $\beta = 0$, $F > \epsilon$.
2 **while** $F > \epsilon$ **do**
3 $\quad \mathbf{p}^* = \arg\max \quad \log_2\left(1 + \frac{\sum_{k=1}^{K} p_k a_k}{\sigma_n^2}\right) - \beta(P_{\text{total}})$;
$\quad\quad$ s.t. (38b), (38c);
4 $\quad F =$
$\quad\quad \log_2\left(1 + \frac{\sum_{k=1}^{K} p_k^* a_k}{\sigma_n^2}\right) - \beta \left(\sum_{k=1}^{K} \nu_k^{-1} p_k^* + c\right)$,
$\quad\quad$ where $c = \sum_{k=1}^{K} P_{\text{UE},k} + P_{\text{BS}} + LP_l$ ;
5 $\quad \beta = \frac{\log_2\left(1 + \frac{\sum_{k=1}^{K} p_k^* a_k}{\sigma_n^2}\right)}{\sum_{k=1}^{K} \nu_k^{-1} p_k^* + c}$;
6 **end**

---

In the following, we propose a low-complexity optimal solution for (39) to reduce the computational load since **Algorithm 3** requires solving (39) many times. Let $F =$ $\log_2\left(1 + \frac{\sum_{k=1}^{K} p_k a_k}{\sigma_n^2}\right) - \beta (P_{\text{total}})$, then the partial derivative of $F$ w.r.t. $p_k$ is calculated as

$$\frac{\partial F}{\partial p_k} = \frac{a_k}{(\ln 2)(\sum_{i=1}^{K} p_i a_i + \sigma_n^2)} - \beta \nu_k^{-1}. \tag{40}$$

By setting $\partial F / \partial p_k = 0$, we get

$$p_k^* = 1/(\beta \nu_k^{-1} \ln 2) - (\sum_{i \neq k} p_i a_i + \sigma_n^2)/a_k. \tag{41}$$

We calculate $p_k^*$ assuming all the remaining $p_i \forall i \neq k$ values are fixed, and this process is repeated until convergence. On this basis, the proposed low-complexity algorithm goes as follows: we first allocate the minimum required power for each user; then, we update the power for the users one by one as illustrated in **Algorithm 4**; this update continues until convergence. Note that the convergence is guaranteed since $F$ increases or remains unchanged after each update, and $F$ has an upper bound. Moreover, the obtained local optimum is also the global optimum since the cost function $F$ in (39) is concave. $p_k^{\min}$ and $p_k^{\max}$ in **Algorithm 4** are calculated for each user according to its boundaries as in the constraints in (38c).

---

**Algorithm 4:** Low complexity and fast iterative solution for solving step 3 in **Algorithm 3**

1 **Initialize:** Set $p_k = p_k^{\min}$, $k = 1, 2, \ldots, K$;
2 **while** 1 **do**
3 $\quad \mathbf{p}_{old} = \mathbf{p}$;
4 $\quad$ **for** $k = 1:K$ **do**
5 $\quad\quad p_k^* = 1/(\beta \nu_k^{-1} \ln 2) - (\sum_{i \neq k} p_i a_i + \sigma_n^2)/a_k$;
6 $\quad\quad$ **if** $p_k^* < p_k^{min}$ **then**
7 $\quad\quad\quad p_k = p_k^{\min}$;
8 $\quad\quad$ **end**
9 $\quad\quad$ **if** $p_k^* > p_k^{max}$ **then**
10 $\quad\quad\quad p_k = p_k^{\max}$;
11 $\quad\quad$ **end**
12 $\quad\quad$ **if** $p_k^{min} < p_k^* < p_k^{max}$ **then**
13 $\quad\quad\quad p_k = p_k^*$;
14 $\quad\quad$ **end**
15 $\quad$ **end**
16 $\quad$ **if** $|\mathbf{p}_{old} - \mathbf{p}| < 10^{-10}$ **then**
17 $\quad\quad$ break while loop;
18 $\quad$ **end**
19 **end**

---

### B. Optimizing the IRS coefficients vector $\mathbf{w}$ given $\mathbf{p}$

Here, we discuss the step of optimizing the IRS coefficients vector $\mathbf{w}$ given the transmit power vector $\mathbf{p}$ for the sake of maximizing the EE of the NOMA system using the alternating maximization algorithm. Both the SDR and manifold optimization based solutions are proposed in this subsection as presented for the total transmit power minimization problem in Sec. III. Unlike the total transmit power minimization problem, in the EE maximization problem, we have a direct objective function of $\mathbf{w}$ to be maximized when optimizing the coefficients vector $\mathbf{w}$ for a given $\mathbf{p}$. When maximizing the objective function in (37a) w.r.t. $\mathbf{w}$, the denominator is neglected because it is considered as a constant as it is not a function of $\mathbf{w}$. Hence, we only focus on the numerator which is the log function. Moreover, the log function is monotonically increasing, hence we can only focus on maximizing its argument. Therefore, when considering the SDR optimization scheme using the same mathematical steps as in Sec. III-B1, the NOMA EE maximization problem in **w** can be written as

$$\max_{\widehat{\mathbf{W}}} \quad \sum_{k=1}^{K} p_k(\text{trace}\{\mathbf{B}_k \widehat{\mathbf{W}}\} + |v_k|^2) \tag{42a}$$

$$\text{s.t.} \quad p_k(\text{trace}\{\mathbf{B}_k \widehat{\mathbf{W}}\} + |v_k|^2) - (2^{R_k^{\min}} - 1)$$
$$\times \left[ \sum_{j=k+1}^{K} p_j(\text{trace}\{\mathbf{B}_j \widehat{\mathbf{W}}\} + |v_j|^2) + \sigma_n^2 \right] \geq 0, \forall k, \tag{42b}$$

$$(17c), (17d),$$

where $\widehat{\mathbf{W}} = \bar{\mathbf{w}} \bar{\mathbf{w}}^H$ and $\bar{\mathbf{w}}$ is defined in (16). Similar to the SDR problem in (17), the problem in (42) is an SDP which is non-convex due to the non-convex rank one constraint in (17d). By dropping (17d), (42) becomes clearly convex which can be directly solved using optimization tools such as CVX.

We can also use the manifold optimization scheme to optimize the IRS coefficients vector, **w**, given **p** when maximizing EE. As discussed in Sec. III-B2, the unit modulus constraints on the elements of **w** restrict the feasible region to the complex circle manifold $\mathcal{M}$. Hence, the EE manifold optimization problem can be written as

$$\max_{\mathbf{w} \in \mathcal{M}} \quad \sum_{k=1}^{K} p_k \left| \mathbf{g}^T \mathbf{H}_k \mathbf{w} + v_k \right|^2 \tag{43a}$$

$$\text{s.t.} \quad C_k(\mathbf{w}) \geq 0, \quad k = 1, 2, \ldots, K. \tag{43b}$$

Following the same procedures as in Sec. III-B2, the problem in (43) is transformed to an unconstrained manifold optimization problem using the exact penalty method. Then, we shall use the same smoothing technique as discussed in Sec. III-B2 to smooth the unconstrained penalized optimization problem. Therefore, the smooth unconstrained NOMA EE manifold optimization problem, in **w** given **p**, can be written as

$$\max_{\mathbf{w} \in \mathcal{M}} \quad \sum_{k=1}^{K} p_k \left| \mathbf{g}^T \mathbf{H}_k \mathbf{w} + v_k \right|^2 + \rho \left( \sum_{k=1}^{K} \mathcal{P}(-C_k(\mathbf{w}), u) \right), \tag{44}$$

where $\mathcal{P}$ is the smoothing function defined in (25). Now, the unconstrained problem in (44) is smooth and differentiable which can be solved using the manifold optimization toolbox as discussed in Sec. III-B2. The parameters $\rho$ and $u$ in (44) are updated in the same way illustrated in **Algorithm 1**.

Surprisingly, as shown in the simulation results in Sec. VI, we found that focusing on expanding the feasible set of the problem in **p** by finding the beamforming coefficients vector **w**, that maximizes the minimum of the constraints $C_k(\mathbf{w})$, is more suitable for the EE maximization problem than maximizing the EE objective function itself. This can be done by the alternation between solving the problem (38) in **p** given **w**, then solving the problem (26) in **w** given **p** in an iterative manner. This approach is counter-intuitive as it optimizes the IRS beamforming vector during **w** phase to expand the feasible set of the power allocation problem in (38) during **p** phase, rather than maximizing the EE itself which is the target of the problem. However, the presented simulation results in Sec. VI show that this approach gives way better EE performance than focusing on maximizing the objective function discussed earlier in this section. The reason behind this behavior is that we optimize the passive beamforming vector, **w**, to expand the feasible region over which the power allocation problem in (38) is optimized during the alternating optimization algorithm. Therefore, when maximizing the EE objective function in the power allocation phase in (38), **Algorithm 4** can search over a larger feasible set to maximize the EE of the system which may increase the value of the obtained EE in the next iteration of the alternation maximization algorithm. The procedures of the proposed CCM based alternation optimization algorithm for the EE maximization problem is summarized in **Algorithm 5**.

---

**Algorithm 5:** EE maximization alternation optimization algorithm

---

1 **Initialize iteration number:** $n = 0$.
2 **Initial feasible point:** $\mathbf{p}_0, \mathbf{W}_0$.
3 **while** $\Xi \geq \epsilon$ **do**
4     Solve (39) in **p**, using **Algorithm 4**, given fixed IRS coefficients matrix $\mathbf{W}_{n-1}$;
5     Solve (26) or (44) in **W**, using **Algorithm 1**, given fixed users' power allocation vector $\mathbf{p}_n$;
6     Calculate $\Xi = EE(\mathbf{p}_n, \mathbf{W}_n) - EE(\mathbf{p}_{n-1}, \mathbf{W}_{n-1})$, where $EE(\mathbf{p}, \mathbf{W})$ is defined in (37a);
7     $n \to n + 1$;
8 **end**
9 **Return:** $\mathbf{p}^*, \mathbf{W}^*$.

---

## V. COMPARISON WITH OPTIMIZED IRS-ASSISTED OMA

In this section, we discuss IRS assisted uplink OMA. Since IRS assisted OMA is much simpler than the NOMA counterpart in terms of optimizing the IRS phases, for fairness-sake, we compare our proposed IRS-NOMA scheme against IRS-OMA to quantify the potential performance gains. Moreover, we examine if the performance of the IRS-OMA scheme can outperform the IRS-NOMA one in any specific cases so we can make a useful conclusion on when one of them is favored over the other. To this end, we study the average transmit power minimization and EE maximization problems for the IRS-OMA scheme, as we did in the NOMA scheme, and provide an optimal solution for both problems.

### A. IRS-assisted OMA power minimization

In this subsection, we present an optimal solution for minimizing the average transmit power of OMA users when assisted by IRS. The sum transmit power is minimized while having minimum uplink rate constraints for the users. In OMA, each user takes a fraction of the time frame to transmit its data. Unlike NOMA, in IRS assisted OMA, each user transmits its data separately within its allocated time fraction, $\alpha_k$, while other users remain idle. $\alpha_k$, is optimized along with the transmit powers, $p_k$, to minimize the average transmit power of the users. This optimization problem can be expressed as

$$\min_{\alpha_k, p_k} \sum_{k=1}^{K} \alpha_k p_k \tag{45a}$$

$$\text{s.t.} \quad p_k \leq P^{\max}, \tag{45b}$$

$$\alpha_k \log_2 \left(1 + \frac{p_k c_k}{\sigma_n^2}\right) \geq R_k^{\min}, \quad k = 1, 2, \ldots, K, \tag{45c}$$

$$\alpha_k \geq 0, \tag{45d}$$

$$\sum_{k=1}^{K} \alpha_k = 1, \tag{45e}$$

where $c_k = |\mathbf{g}^T \mathbf{W}_k \mathbf{h}_k + v_k|^2$ and $P^{\max}$ is the maximum allowable uplink transmit power for the users. As long as the users send their data separately during their allocated time, the IRS adjusts its phase shifts, $\mathbf{W}_k$, independently of each user according to the corresponding channel vector, $\mathbf{h}_k$, as each user experiences an *interference-free* channel. The IRS phase shifts are chosen so that all the elements of the cascaded user-IRS-BS channel, $h_{ki} w_i g_i \forall i$, are aligned to have the same phase which is the phase of the direct channel $v_k$, i.e., $\theta_i = \theta_{v_k} - \theta_{g_i} - \theta_{h_{ki}}$. To minimize the average transmit power of the OMA users, the minimum rate constraints in (45c) is satisfied with equality to reduce $p_k$ as much as possible. When the constraints in (45c) are satisfied with equality, the power of each user $p_k$ can be expressed in a closed-form as

$$p_k = \frac{\sigma_n^2}{c_k} \left(2^{R_k^{\min}/\alpha_k} - 1\right). \tag{46}$$

By eliminating $p_k$ from (45a), (45) can be reduced to

$$\min_{\alpha_k} \sum_{k=1}^{K} \frac{\sigma_n^2 \alpha_k}{c_k} \left(2^{R_k^{\min}/\alpha_k} - 1\right) \tag{47}$$

$$\text{s.t.} \quad (45b), (45d), (45e).$$

The feasible set of the optimization problem (47) is clearly convex since the constraints (45d) and (45e) are affine in $\alpha_k$. Now, we need to check the convexity of the objective function in (47). The objective function is separable in $\alpha_k$'s, i.e. it can be written as a sum of separate terms in $\alpha_1, \alpha_2, \ldots, \alpha_K$. Therefore, the Hessian matrix of (47) is diagonal as $\frac{\partial f_{oma}}{\partial \alpha_k \partial \alpha_j} = 0$, $k \neq j$, where $f_{oma}$ is the objective function in (47). Since the Hessian matrix is diagonal, its eigenvalues are the diagonal elements themselves, which can be calculated as

$$\frac{\partial^2 f_{oma}}{\partial \alpha_k^2} = \frac{\sigma_n^2}{c_k} \frac{((\ln 2) R_k^{\min})^2}{\alpha_k^3} 2^{(R_k^{\min}/\alpha_k)}, \tag{48}$$

where $\frac{\partial^2 f_{oma}}{\partial \alpha_k^2}$ is the second partial derivative of $f_{oma}$ w.r.t. $\alpha_k$. Clearly, the second derivatives in (48) are always positive over the feasible set of (47) since $\alpha_k$'s are non-negative. Consequently, (47) is strictly convex since it has a convex feasible set as well as a convex objective function, then it can be efficiently solved using CVX or MATLAB's fmincon function in the optimization toolbox.

*B. IRS-assisted OMA EE maximization*

In this subsection, we present the EE maximization problem of the OMA system and propose an optimal solution for it. The EE maximization problem is solved under minimum uplink data rate constraints for the users. The EE maximization problem of the OMA system can be formulated as

$$\max_{\alpha_k, p_k} \frac{\sum_{k=1}^{K} \alpha_k \log_2 \left(1 + \frac{p_k c_k}{\sigma_n^2}\right)}{\sum_{k=1}^{K} \alpha_k p_k} \tag{49}$$

$$\text{s.t.} \quad (45b), (45c), (45d), (45e).$$

Assuming that $c_1$ is greater than the values of $c_k$ of other users, then we focus on allocating all the excess power to user 1, i.e., increasing $p_1$, while the powers of the other users are chosen to satisfy their minimum rate constraints with equality. Therefore, the EE maximization problem reduces to

$$\max_{\alpha_k, p_1} \frac{\alpha_1 \log_2 \left(1 + \frac{p_1 c_1}{\sigma_n^2}\right) + \sum_{k=2}^{K} R_k^{\min}}{\alpha_1 p_1 + \sum_{k=2}^{K} \alpha_k \left[2^{R_k^{\min}/\alpha_k} - 1\right] \frac{\sigma_n^2}{c_k}} \tag{50a}$$

$$\text{s.t.} \quad \alpha_1 \log_2 \left(1 + \frac{p_1 c_1}{\sigma_n^2}\right) \geq R_1^{\min}, \tag{50b}$$

$$(45b), (45d), (45e).$$

Problem (50) is solved using alternating maximization over $p_1$ and $\alpha_k$ iteratively until convergence. It can be readily proved that the problem in (50) is concave w.r.t. $p_1$. However, (50) is quasi-concave in $\alpha_k$ since the numerator is affine and the denominator is convex w.r.t. $\alpha_k$ as proved in the previous subsection. Therefore, the Dinkelbach algorithm discussed in Sec. IV-A can be used to solve (50) w.r.t. $\alpha_k$.

## VI. SIMULATION RESULTS

In this section, we provide extensive simulation results to validate the proposed optimization algorithms for IRS assisted uplink NOMA systems. The proposed solutions are compared against the IRS assisted OMA under different scenarios. For each optimization problem, we compare the performance of the proposed manifold algorithm against three SDP-based benchmarks which are SDR, SDP-DC and SROCR. Under each scenario, we present the results in terms of the total transmit power, EE, and the achievable uplink sum rate of the proposed solution. Comparing these three system performance measures of the total transmit power minimization problem against those of the EE maximization problem reveals very useful insights which relate the two problems. The two problems are solved and compared against each other under two scenarios; the first is when low minimum target uplink rates are required for the users, and the second is when high minimum rate constraints are required for the users. The number of users, $K$, in the NOMA cluster is assumed 3.

For fair comparison, similar simulation parameters to those used in [6] and [36] have been also considered in this section, which can be summarized as follows. The distance between BS and IRS is assumed $d_{IB} = 75$ m. The distances between the 3 users and IRS are $\{10, 20, 40\}$ m, whereas the distances between the users and BS are $\{30, 50, 200\}$ m. The path loss at the reference distance in (3) is $\eta_0 = 10^{-3}$, and the path loss exponents for the direct links (users to BS), IRS to users links and BS to IRS link are assumed in this simulation setup as $\alpha_{BU} = 5.5$, $\alpha_{IU} = 2.2$ and $\alpha_{BI} = 2.2$, respectively. The noise power $\sigma_n^2$ is set to be $-114$ dBm, and the Rician factors in (1) and (2) are set to be $K_{IB} = K_{UI} = 2.2$. The maximum transmit power constraint, $P^{\max}$, in (37b) is set to 10 dBm in case of low minimum rate constraints, and 25 dBm in case of high minimum rate constraints for the users.

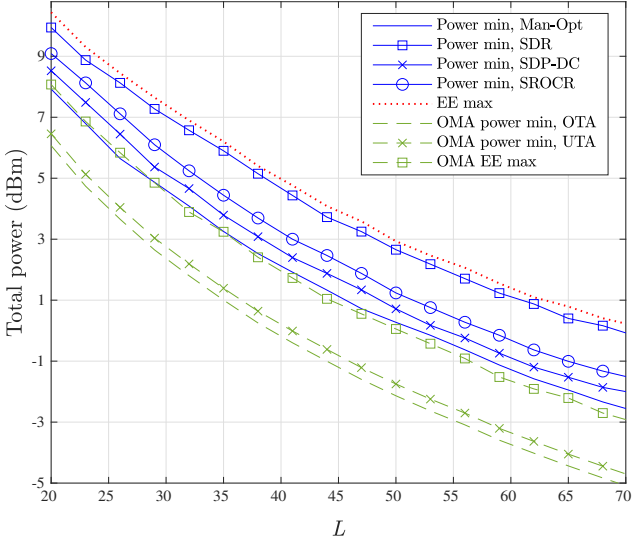

Figure 2: Sum power vs. No. of reflectors $L$, $R_k^{\min}$=0.2 bits/sec/Hz

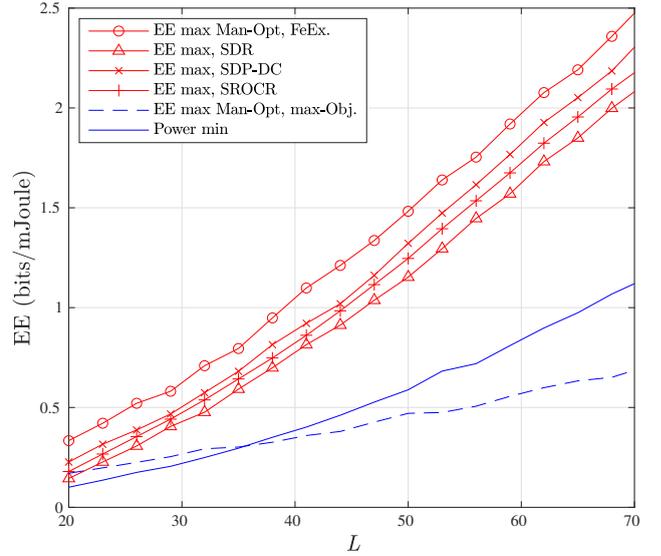

Figure 3: EE vs. No. of reflectors $L$, $R_k^{\min}$=0.2 bits/sec/Hz

### A. NOMA vs. OMA for low minimum rate constraints

In Fig. 2, we present the total transmission power of the users versus the number of reflecting elements at IRS for different schemes. Low QoS constraints at the users are assumed in Fig. 2 where the minimum target achievable rates at the users are set at 0.2 bits/sec/Hz. We compare the performance of the proposed scheme against SDR, SDP-DC, SROCR and the IRS assisted OMA scheme. The results of the IRS-OMA scenario are included when optimal time allocation (OTA) and uniform time allocation (UTA) are adopted. The figure shows that our proposed optimization scheme outperforms SDR, SDP-DC and SROCR, although the complexity of the three benchmarks is much higher than the complexity of the proposed algorithm. Fig. 2 also shows that the IRS assisted OMA scheme outperforms the NOMA scheme for the case of low minimum target rate requirements for the users, hence, the IRS-OMA scheme is preferable in this case. The graph also compares the total transmit power of the users resulting from solving the sum power minimization problem against the total power resulting from solving the EE maximization problem for both NOMA and OMA cases. Clearly, and as expected, the graph shows that the sum power curve of the EE maximization problem is higher than the total power resulting from the power minimization problem. The rationale behind this is that the EE objective function, in (38a) for the NOMA case or in (49) for the OMA case, is pseudo concave w.r.t. the transmit power vector $\mathbf{p}$. Hence, when the target uplink rate constraints are low, the minimum transmit powers of the users must be low too, which makes the system's EE far from the maximum EE point of the pseudo concave objective function. Therefore, in the case of low rate constraints, moving towards the maximum EE point requires increasing the transmit powers of the users above the minimum. That is why the power consumption in the case of NOMA/OMA EE maximization is higher than that in the case of total transmit power minimization.

Fig. 3 shows the performance of the proposed EE maximization solutions using the proposed CCM based alternation maximization algorithm in **Algorithm 5** discussed in Sec. IV. The IRS assisted NOMA EE maximization problem is solved by optimizing $\mathbf{p}$ given fixed $\mathbf{w}$ using the proposed low complexity **Algorithm 4**, then optimizing $\mathbf{w}$ given fixed $\mathbf{p}$ alternately using the alternation maximization algorithm. When solving for $\mathbf{w}$ given $\mathbf{p}$, we compare the proposed manifold optimization based feasibility expansion scheme, discussed in Sec. IV-B, against the SDR, SDP-DC and SROCR benchmarks. The figure shows that the proposed manifold optimization algorithm, when applying the feasibility expansion (FeEx.) mechanism discussed in Sec. IV-B, outperforms the three benchmarks, although it has much lower complexity. The figure also confirms that the proposed feasibility expansion mechanism provides far better performance by solving (26) than maximizing the EE cost function using (44). The proposed CCM based feasibility expansion (FeEx.) mechanism in **Algorithm 5** outperforms the existing SDP based counterparts because it optimizes the original optimization vector, $\mathbf{w}$, on the original feasibility region as in (26). On the other hand, SDP based schemes like SDR, as in (42), broaden the original feasibility region, encompassing a larger set, $\mathbf{W}$, which is the outer product of the original optimization vector. When the optimal solution obtained on the larger set, in the SDP algorithms, is projected back on the original feasible region through Cholesky decomposition, the results may deviate from the optimal solution to the original problem which causes performance reduction compared to the proposed CCM-based approach. Moreover, since the problem (37) focuses on maximizing the EE of the system, it is expected to see that the EE resulted from (37) is higher than the EE resulted from solving the sum power minimization problem in (8). This result is because problem (8) only focuses on minimizing the transmit powers of the users without paying any attention to the EE metric. On the other hand, in the EE maximization problem (37), the users are permitted to increase their transmit powers, and hence, increase their uplink data rates accordingly above the minimum target rates to increase EE.

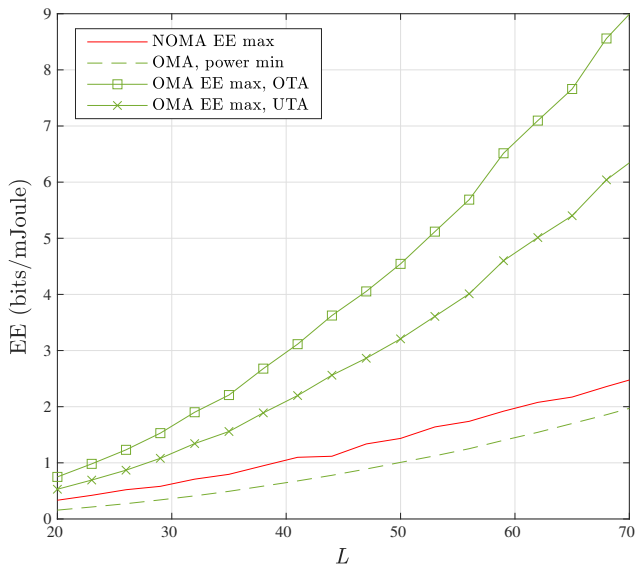

Figure 4: Comparing OMA and NOMA EE, $R_k^{\min}$=0.2 bits/sec/Hz

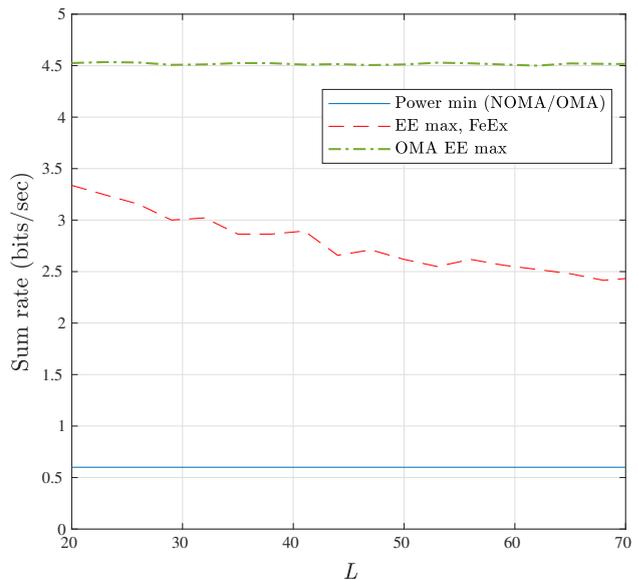

Figure 5: Sum rate vs. No. of reflectors $L$, $R_k^{\min}$=0.2 bits/sec/Hz

Fig. 4 compares the EE of the IRS assisted OMA scheme against the EE of the NOMA counterpart for the case of low minimum uplink rate constraints at the users. The graph shows that the resultant EE of the OMA scheme largely beats the EE of the NOMA scheme in the case of low uplink QoS requirements at the users. Fig. 4 also shows that optimizing the time allocations of the IRS assisted OMA scheme, by solving problem (49) in Sec. V-B, results in higher EE than the regular OMA scheme with equal time fractions among the users. The figure also compares the resultant EE of the OMA scheme when the objective is minimizing the total transmit power against the EE resulted from maximizing the EE of the system. Intuitively, the resultant EE of maximizing the OMA system's EE is higher than the resultant EE of minimizing the OMA total transmit power. This is because the OMA EE objective function in (49) is pseudo concave w.r.t. the transmit power vector $\mathbf{p}$. When the target uplink rate constraints are low, the user transmit at low power levels in the case of total transmit power minimization, which makes them far behind the maximum EE point. Therefore, in the case of OMA EE maximization, the users increase their transmit powers moving towards the maximum EE point of the OMA system. However, in the case of minimizing the total transmit power, we only focus on minimizing the transmit power as much as possible by satisfying the minimum target uplink rates with equality while neglecting the EE aspect.

In Fig. 5, we depict the attainable sum rates of users within the system, attained by solving both the sum power minimization and the EE maximization problems in (8) and (37), and their OMA versions (45) and (49), respectively. When addressing the total transmit power minimization problem in a NOMA/OMA system, each user precisely meets the required minimum uplink rate with equality. This outcome aligns with intuition, as the primary objective is to minimize the transmit powers of users as much as possible in this scenario. In contrast, when tackling the EE maximization problem in a NOMA/OMA system, users have the flexibility to transmit at higher power levels, achieving elevated uplink data rates to maximize the overall EE of the system. As illustrated in Fig. 2 and discussed in the paper, the EE cost functions for NOMA and OMA, represented by equations (38a) and (49), respectively, exhibit pseudo-concavity w.r.t. the transmit power vector, $\mathbf{p}$. Consequently, when the constraints on target uplink rates are low, users increase their transmit power, moving towards the system's maximum EE point. Consequently, the achievable uplink data rates of users surpass the target minimum rates when addressing the EE maximization problem for both NOMA and OMA systems. This behavior underscores the adaptive nature of users in response to EE maximization, allowing them to dynamically adjust their transmit power levels to optimize system-wide EE.

Fig. 6 compares the convergence behaviours of the alternation optimization algorithm using the proposed manifold optimization technique against the different SDP-based benchmarks that we use in this paper. The figure shows the convergence curves of both the sum power minimization and the EE maximization problems. The graph shows that the proposed alternation optimization based manifold optimization algorithm converges after few iterations, and faster than the shown benchmarks, when solving the two considered optimization problems. These results confirm the convergence of the proposed alternation optimization algorithm in Sec. III-C.

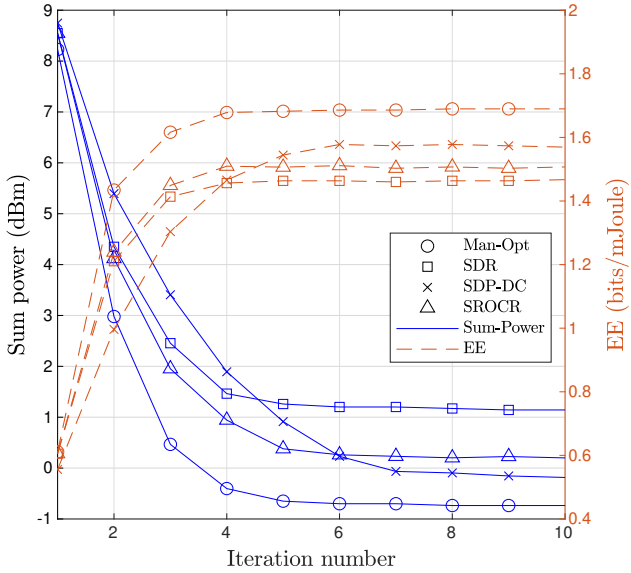

Figure 6: Convergence of sum power and EE optimization using proposed manifold and benchmarks, $R_k^{\min}$=0.2 bits/sec/Hz

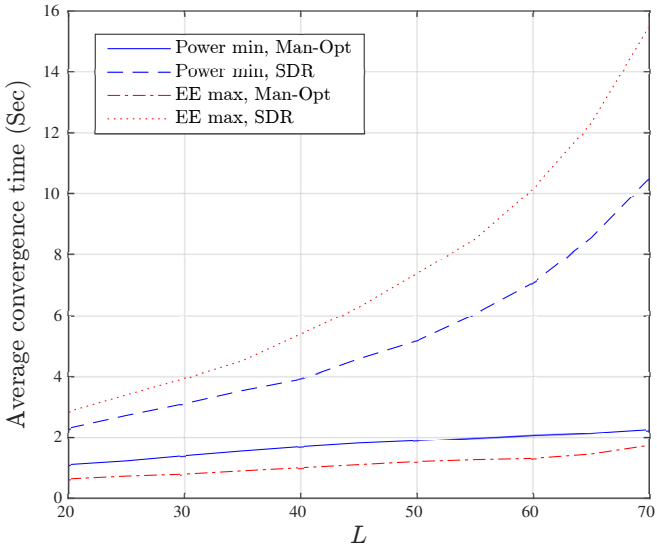

Figure 7: Time complexity of sum power and EE optimization using proposed manifold vs. SDR, $R_k^{\min}$=2 bits/sec/Hz

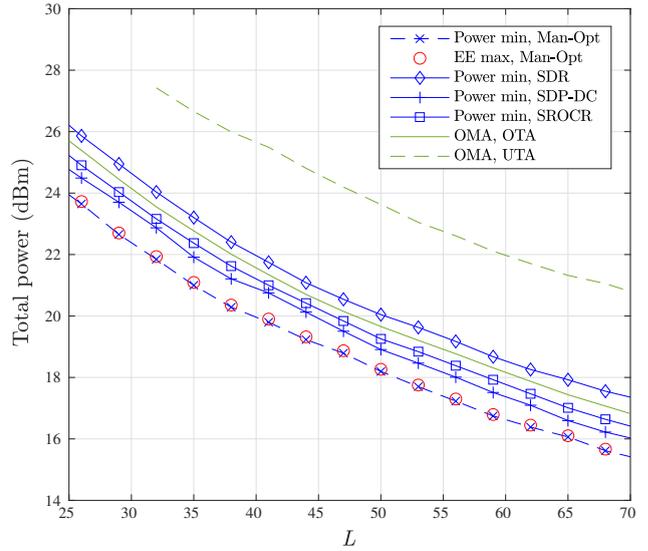

Figure 8: Sum power vs. No. of reflectors $L$, $R_k^{\min}$=2.5 bits/sec/Hz

Fig. 7 compares the time complexity of the proposed CCM algorithms in **Algorithm 2** and **Algorithm 5** against the SDR benchmark for solving the sum power minimization and the EE maximization problems, respectively. We plot the average run time per iteration for each optimization algorithm against the number of IRS elements. Clearly, the figure shows that the average convergence run time of the proposed CCM based optimization algorithm is far less than the average run time of the SDR benchmarks for both the sum power minimization and the EE maximization problems. These results yield a huge time complexity reduction for our proposed CCM based algorithm against the SDR-based benchmarks like SDR, which confirms the presented complexity analysis in Sec. III-D.

### B. NOMA vs. OMA for high minimum rate constraints

In the next figures, we show the performance of the proposed solutions to both the total transmit power minimization problem and the EE maximization problem when the minimum target uplink rates are high. We assume that the target minimum transmission rate for each user is 2.5 bits/sec/Hz. We will show how the two optimization problems are related when the target transmission rates of the users are high.

Fig. 8 presents the relationship between the total uplink transmit power of users and the number of reflecting elements for various schemes. The proposed manifold scheme is compared against benchmarks such as SDR, SDP-DC, and SROCR, as well as an IRS-assisted OMA scheme. Similar to the scenario with low rate constraints, the manifold optimization-based approach in (26) outperforms the three benchmarks in the high rate constraints case too, all while maintaining significantly lower computational complexity. In contrast to the low rate constraints scenario, Fig. 8 reveals that the IRS-NOMA scheme in (8) surpasses the optimized IRS-OMA scheme in (45) when there are high minimum rate requirements for users. This implies that IRS-NOMA is more favorable than IRS-OMA when dealing with stringent minimum target rate constraints. Additionally, Fig. 8 compares the total transmit power of the users resulting from solving the sum power minimization problem against the total power from solving the EE maximization problem for both NOMA and OMA cases. Interestingly, in cases with high rate constraints, the sum power curve resulting from solving the EE maximization problem (37) perfectly coincides with the sum power curve from solving the total transmit power minimization problem (8). This behavior is explained by the pseudo-concave nature of the NOMA EE objective function in equation (38a) w.r.t. the transmit power vector **p**. Consequently, when the target uplink rate constraints are high, the minimum transmit powers of users must also be high, exceeding the maximum EE point. Therefore, in scenarios with high rate constraints, moving towards the maximum EE point requires decreasing the transmit powers of the users. Hence, maximizing EE directly leads to minimizing the total transmit power of users, resulting in the overlapping of both solutions.

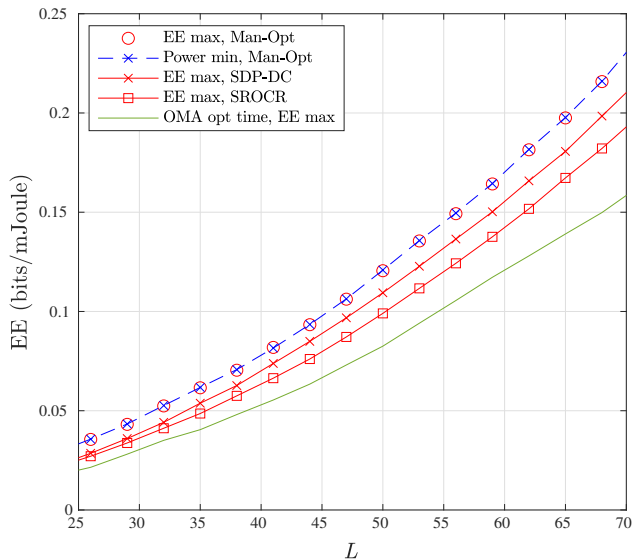

Figure 9: EE vs. No. of reflectors $L$, $R_k^{\min}$=2.5 bits/sec/Hz

Fig. 9 shows a performance comparison between the proposed EE maximization solutions using the alternation maximization scheme discussed in Sec. IV for the case of high uplink rate requirements at the users. Similar to the case of low rate constraints at the users, the figure shows the performance gain of the obtained EE using the proposed manifold optimization over the SDR, SDP-DC and SROCR benchmarks. It is clear from Fig. 9 that the EE resulted from minimizing the total transmit power by solving (8) is exactly the same as the EE resulted from maximizing the EE of the system by solving (37). This agrees and supports the results and discussion of Fig. 8 since we found that the EE maximization problem co-insides with the total transmit power minimization. Therefore, the two problems lead to the same EE values in the case of high uplink rate requirements.

## VII. CONCLUSIONS AND FUTURE WORK

In this article, we proposed efficient solutions to the sum power minimization and the EE maximization problems for IRS assisted uplink NOMA networks. The two problems are solved by jointly optimizing the users' uplink transmit powers and the IRS passive beamforming coefficients using alternation optimization based algorithms. During the passive beamforming sub-problem, we showed that the proposed manifold optimization solutions have significant performance gains and are far superior to the SDR benchmark solutions. Additionally, we compared our solutions for the IRS assisted NOMA system with the IRS assisted OMA counterpart.

The simulation results indicate that NOMA exhibits superior performance over OMA when the users' uplink data rate requirements are high. Conversely, OMA is more favorable when the rate requirements are low. This behaviour can be attributed to the fact that there is an optimal closed-form solution for the phase shifts of IRS elements with OMA signalling, which provides some performance gain over IRS-NOMA. This gain might overwhelm the gain obtained by IRS-NOMA when the minimum data rate requirements are low because the spectral efficiency gained by NOMA signalling is negligible under these operating conditions as the achievable rate of the near user is much higher than the other users. On the other hand, when the minimum rate requirements are high, the achievable rates for all users will be comparable, which implies that the system operates in the mid capacity region in which NOMA signalling has a superior gain. This conclusion actually does not undermine the benefits of IRS-NOMA since it can be used in applications that require relatively high data rates such as UAV communications and IoT data collection points. On the other hand, IRS-OMA is useful for low rate applications such as wireless sensor networks (WSN).

In future work, it is worth investigating the performance of the developed low complexity CCM based algorithm to optimize uplink NOMA systems with multiple antennas at the BS. It will be interesting to alternatively optimize the IRS phase shifts, using the CCM algorithm, the transmit powers of the users, and the NOMA receive beamforming vector.